\documentclass[referee]{raa}            % referee version: for submission
\newcommand{\target}{EPIC 202843107 }
\newcommand{\uhz}{ $\mu$Hz }
\newcommand{\kep}{{\it Kepler }}
\newcommand{\ktwo}{{\it K2 }}

\usepackage{graphicx,times}             
\usepackage{natbib}
\usepackage{amssymb,amsmath}
\bibpunct{(}{)}{;}{a}{}{,}

\usepackage[a4paper=true,dvipdfm=true,pagebackref=true]{hyperref}
\hypersetup{colorlinks = true, linkcolor = green, anchorcolor = red, citecolor = blue, filecolor = red, pagecolor = red, urlcolor = red}

\begin{document}

   \title{\target: A Close Eclipsing Binary Containing a $\delta$ Scuti Variable
}
   \volnopage{Vol.0 (20xx) No.0, 000--000}     
   \setcounter{page}{1}          

   \author{Jian-Wen Ou
      \inst{}
   \and Ming Yang $^{\dagger}$
      \inst{}
   \and Ji-Lin Zhou $^{\ddagger}$
      \inst{}
   }
   \institute{School of Astronomy and Space Science \& Key Laboratory of Modern Astronomy and Astrophysics in Ministry of Education, Nanjing University, Nanjing 210093, China;\\
    $^{\dagger}${\it ming.yang@nju.edu.cn},
    $^{\ddagger}${\it zhoujl@nju.edu.cn}\\
%   \institute{National Astronomical Observatories, Chinese Academy of Sciences, Beijing 100012, China; {\it aiying@bao.ac.cn}\\
\vs\no
   {\small Received~~20xx month day; accepted~~20xx~~month day}}

\abstract{ This paper reports on the discovery that an eclipsing binary system, \target, has a $\delta$ Scuti variable component. The phased light curve from \kep space telescope presents a detached configuration. The binary modelling indicates that the two component stars have almost the same radius and may have experienced orbital circularization. Frequency analyses are performed for the residual light curve after subtracting the binary variations. The frequency spectrum reveals that one component star is a $\delta$ Scuti variable. A large frequency separation is cross-identified with the histogram graph, the Fourier transform, and the echelle diagram method. The mean density of the $\delta$ Scuti component is estimated to be $0.09$ g$\cdot$cm$^{-3}$ based on the large separation and density relation. Systems like \target are helpful to study the stellar evolution and physical state for binary stars.
}

   \authorrunning{J.-W. Ou, M. Yang \& J.-L. Zhou }            %author_head in even pages
   \titlerunning{\target: A Close Eclipsing Binary Containing a $\delta$ Scuti Variable}  % title_head in odd pages

   \maketitle
%% The author head (on even pages) and the title head (on odd pages) will be
%% automatically extracted from \author{} and \title{}. Whenever the title is too long,
%% you will be asked to supply a shorter one by inserting either \authorrunning{} or
%% \titlerunning{} before \maketitle. Anyway, you can specify your own heads.
%%
%%
%% Note: In the following text body of your manuscript, please note several differences from
%%       other major journals:
%% (1) \subsection{Please Capitalize the First Letter of Each Notional Word in Subsection Title}
%% (2) Please Capitalize the First Letter of Each Notional Word in all tables' captions

%
%________________________________________________ sections below
%
%%%%%%%%%%%%%%%%%%%%%%
%%%%%%%% Section 1  %%%%%%%%
%%%%%%%%%%%%%%%%%%%%%%
\section{Introduction} 
\label{sec1}
The delta Scuti ($\delta$ Sct) variables represent a class of asteroseismic objects which locate at the bottom of the Cepheid instability strip on the Hertzsprung-Russell (HR) diagram. Their typical brightness fluctuations range from 0.003 to 0.9 magnitudes with pulsating periods between about 0.3 and 8 hours \citep{Aer10}. Their intrinsic pulsations is driven by the $\kappa$ mechanism, which originates from the partial ionization zone of helium.

The eclipsing binary with a $\delta$ Sct component is valuable for testing the stellar theoretical models and the interactions between stars. For example, some researchers \citep{Lia12, Kah17, Lia17} reveal the correlations between the pulsation frequency and stellar fundamental characteristics. Furthermore, a scaling relation between the mean stellar density and the pulsation frequency has been derived based on the observations of these systems \citep{Gar15}.

Continuous photometric data with high precision were acquired thanks to the NASA's \kep space telescope \citep{Bor08}. The data of \kep are quite valuable for the study of time-domain astronomy, e.g. investigating eclipsing binaries \citep{Prv11, Bor16, Kir16} and $\delta$ Sct stars \citep{Bal11, Guo16, Bow18}. The \ktwo mission \citep{How14} took the place of the \kep mission when two of the four reaction wheels failed in May 2013. The \ktwo mission observed the fields along the ecliptic plane. Each \ktwo campaign lasted for about 80 days with a photometric precision $\sim$80 ppm over 6 hours \citep{How14}. In order to distinguish each target, a new input catalogue, the Ecliptic Plane Input Catalogue (EPIC), has been used for the \ktwo mission.

\target (RA=16:22:11.469, Dec=-28:09:42.56) falls in the second campaign field of the \ktwo mission. The observation of \target lasted for 77.29 days from August 23 to November 10, 2014. \target was first identified as an exoplanet candidate by \cite{Van14}. However, \cite{Arm16} classified \target as an eclipsing binary using the machine learning technique. \cite{Bar16} also pointed out its binary nature by analysing the de-correlated flux-position of the photometric pixel file. 

In this paper, we discover that \target is a short-period eclipsing binary system with a $\delta$ Sct variable. The structure of this paper is as follows. Section \ref{sec2} describes the observational data. Section \ref{sec3} solves the light curve with binary modelling and obtains the binary parameters. Section \ref{sec4} analyses the stellar pulsations and confirms that one component star is a $\delta$ Sct variable. Regular frequency spacing is described in section \ref{sec4.1}, and the mean density of the $\delta$ Sct component is estimated in section \ref{sec4.2}. Finally, section \ref{sec5} presents the results and conclusions.

%%%%%%%%%%%%%%%%%%%%%%
%%%%%%%% Section 2  %%%%%%%%
%%%%%%%%%%%%%%%%%%%%%%
\section{Observations}
\label{sec2}
\begin{figure}[ht!]
\begin{center}
\includegraphics[width=1.0\textwidth]{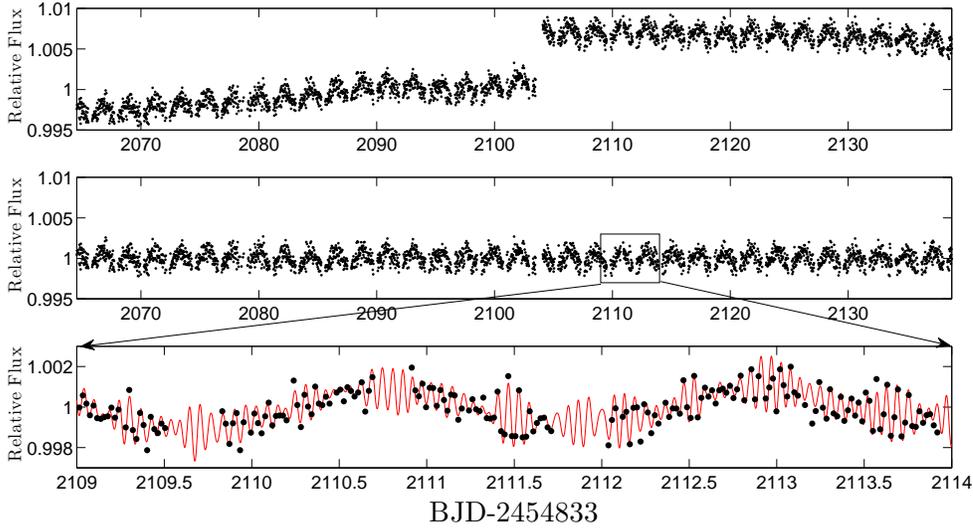}
\caption{Original (top panel) and detrended (middle panel) light curve of \target. Eclipses are not shown for a better view. The bottom panel presents a short section of the detrended light curve marked by the inset box in the middle panel. Solid line illustrates the pulsations superposed on the binary effects.
}
\label{Fig1}
\end{center}
\end{figure}

The data of \target come from the Mikulski Archive for Space Telescopes (MAST; https://archive.stsci.edu/). The raw data contain some instrumental noises, which should be removed before data analysis. Three types of data processing techniques have been applied. \cite{Van14} corrected the \ktwo data using the centroid position of the stellar images. \cite{Arm16} adopted a similar method and developed an analytical procedure specifically for variable stars. \cite{Lug16} applied pixel level de-correlation to remove instrumental noise. However, the long-term trends, which means  continuous rise or fall, still exist in the light curves. The trends can be removed by a low-order polynomial fitting. We adopt a third-order polynomial function to remove the trends in the data of \cite{Lug16}. There is an obvious jump around 2104 as shown in the top panel of Figure \ref{Fig1}. Thus the light curve of \target are divided into two segments. For each segment, we only use the out-of-eclipse part to fit the long-term trend. Then the fitting function is extended to the global segment (including eclipses). The detrended light curve is shown in the bottom panel of Figure \ref{Fig1}.

\section{Light curve solutions}
\label{sec3}
In order to determine reliable parameters for this eclipsing binary system, we analyse the light curve of \target using the JKTEBOP code \citep{Sou04, Sou05, Sou09}. JKTEBOP is a fast and convenient package for modelling light curves of eclipsing binary stars and transiting exoplanets. It is widely used for fitting \kep data \citep{Bog15, Gau16, Man16}. This code can find appropriate parameters for a global minimum that best reproduces the observational data. The fitting parameters include the surface brightness ratio $S_2/S_1$, the ratio of the radii $R_2/R_1$, the sum of radii over the semi-major axis $(R_1+R_2)/a$, the orbital inclination $i$, the eccentricity $e$, and the orbital period $P_{\rm orb}$. 

Firstly, the input values of the fitting parameters are acquired based on the morphology of the eclipsing light curve. The input orbital period is estimated to be $4.398$ days from Fourier analysis. The eccentricity and the periastron longitude degree are set as 0 according to the equal distance between primary and secondary eclipses. The fractional radius is estimated around 1 depending on the widths of the primary and secondary eclipses. The input inclination is set around $90^\circ$. Other parameters are assigned as their typical values. Then, JKTEBOP fits the light curve with these input values using Levenberg-Marquardt minimisation \citep{Sou04} and produces an output parameter file. During the fitting process, JKTEBOP can automatically generate different sets of initial values based on the input values to acquire global minimum. Finally, reliable uncertainties are estimated with Monte Carlo simulations.

\begin{figure}[ht!]
\begin{center}
\includegraphics[width=1.0\textwidth]{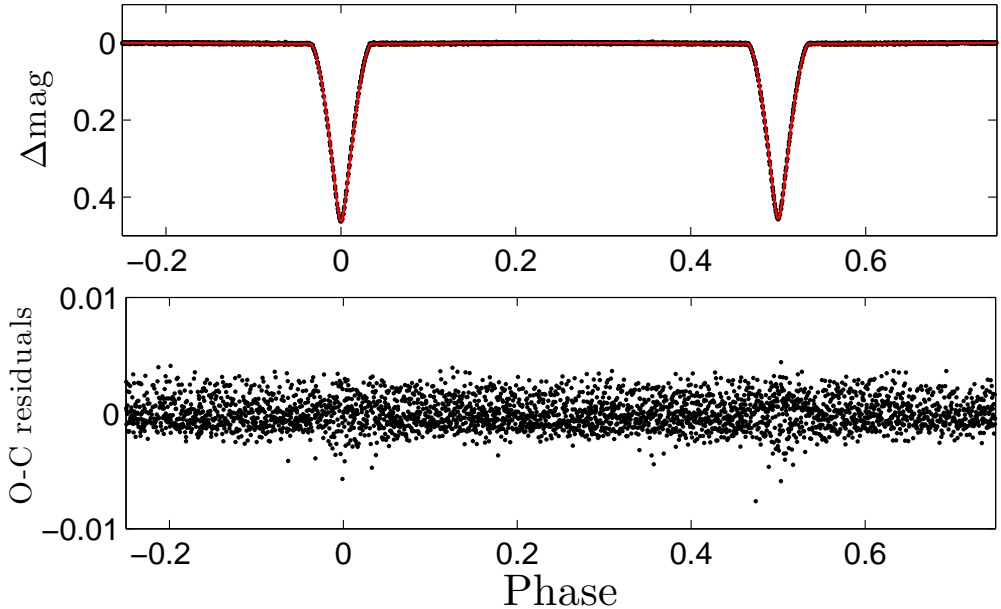}
% \vspace{2mm}
\caption{Binary modelling of \target. The top panel is the observed light curve (black dots) with the best-fit result (red solid line). The bottom panel is the O-C residuals to be used for pulsation analysis.}
\label{Fig2}
\end{center}
\end{figure}

The best-fit parameters of the binary modelling are listed in Table \ref{Tab1}, with the modelled light curve and residuals presented in Figure \ref{Fig2}. The residuals all concentrate around zero. Therefore, the fitting results are reliable. We can see that the two component stars have almost identical radius. The orbital eccentricity is very small, which means this system may have experienced the process of circularization. 
		
%%%%%%%%%%%%%%%% Table 1 %%%%%%%%%%%%%%%%%%
\begin{table*}
\caption{Light curve solution and orbital parameters of \target. }              % title of Table
\centering                                      % used for centering table
%\begin{tabular}{c c l c c c l c l c c }        
\begin{tabular}{c c c c}          % centered columns (11 columns)
\hline\hline                        % inserts double horizontal lines
Parameter & Value & Parameter & Value \\     % table heading
\hline
$S_2/S_1$ & $1.0048^{+0.0153}_{-0.0141}$ & $i $ ($^{\circ}$) & $88.8000^{+0.0405}_{-0.0381}$\\
$R_2/R_1$ & $1.0159^{+0.0152}_{-0.0145}$ & $ e $ & $0.0025^{+0.0024}_{-0.0022}$\\
$(R_1+R_2)/a$ & $0.2167^{+0.0005}_{-0.0005}$ & $P_{\rm orb}$ (days) & $4.397793^{+0.000004}_{-0.000004}$ \\
\hline   \hline                        % inserts double horizontal lines
\end{tabular}\\
\label{Tab1}
\vspace{3mm}
\tablecomments{0.6\textwidth}{Parameters are surface brightness ratio ($S_2/S_1$), ratio of the radii ($R_2/R_1$), the sum of radii over the semi-major axis ($(R_1+R_2)/a$), orbital inclination ($i$), eccentricity ($e$), and orbital period ($P_{\rm orb}$).}
\end{table*}

\section{Frequency analysis}
\label{sec4}
\begin{figure}[ht!]
\begin{center}
\includegraphics[width=1.0\textwidth]{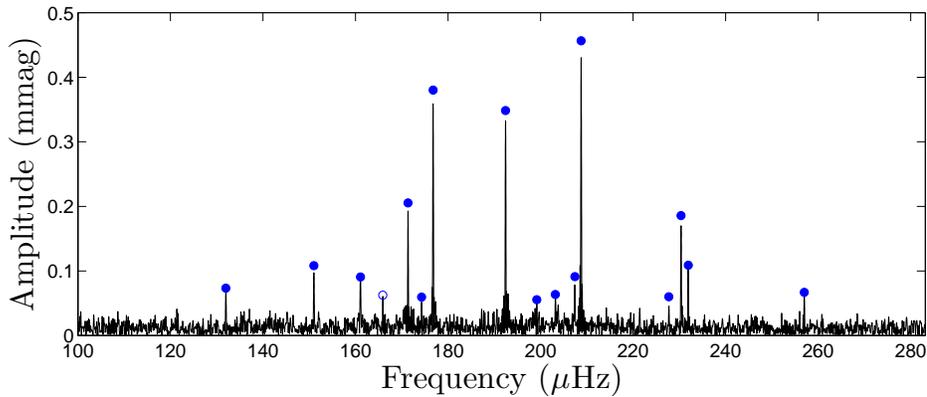}
\caption{The frequency spectrum of the residual light curve after subtracting the binary model. Filled circles represent the frequencies in Table \ref{Tab2}, and the open circle is the 63-order harmonic of the orbital frequency.} 
\label{Fig3}
\end{center}
\end{figure}

Stellar pulsations of \target can be obtained by subtracting the binary model from the detrended light curve. In order to investigate these variations in detail, frequency spectrum analysis is performed based on the residual light curve. The Period04 algorithm \citep{Len05} is adopted to acquire the frequencies, amplitudes, and signal-to-noise ratios (S/N). To ensure the robustness of the extracted pulsation frequencies, we only retain the frequencies with a signal-to-noise ratio larger than 4. As a result, a total of 16 frequencies are found and are listed in Table \ref{Tab2}. The frequency spectrum is shown in Figure \ref{Fig3}. The main pulsations lie in the frequency region between 120\uhz and 270\uhz, with the dominant peak at $f_{12}=208.769$\uhz. According to the pulsation periods, amplitudes, and the profile, we confirm that \target has a $\delta$ Sct component star. 

Mode identification -- -- the determining of pulsation mode values $l$ -- -- is difficult in the analysis of pulsating variable stars. Generally, pulsation modes can be identified by comparing the phase lags between the light curves at different wavelength \citep{Bre99,Pap18}. Another technique is to analyse the spectroscopic equivalent width and intensity because different mode values cause different variations \citep{Bed96,Bal00}. Although \target do not have multi-colour photometry and spectroscopic observation, there are still some useful information that can be derived from the analysis of the frequency spectrum. Studies have found that there exists large frequencies separations for some $\delta$ Sct stars. The large separation presents a regular spacing and is related with the physical state of the star. In the following we will describe the regular spacing study of \target.

%Table 2        
\begin{table*}
\caption{Extracted frequencies of \target.} 
\centering   % used for centering table
\begin{tabular}{c c c c c c}          % centered columns (11 columns)
\hline\hline                        % inserts double horizontal lines
ID & Frequency (\uhz) & Amplitude (mmag) & Phase ($2\pi$/rad) & S/N & Note \\
\hline
$f_{1}$	&	131.982	&	0.073	&	0.071	&	5.72	&		\\
$f_{2}$	&	151.001	&	0.108	&	0.589	&	8.83	&		\\
$f_{3}$	&	161.090	&	0.091	&	0.132	&	7.57	&		\\
$f_{4}$	&	165.906	&	0.062	&	0.821	&	5.09	&	$63f_{\rm orb}$	\\
$f_{5}$	&	171.361	&	0.205	&	0.387	&	16.72	&		\\
$f_{6}$	&	174.285	&	0.059	&	0.809	&	4.87	&		\\
$f_{7}$	&	176.767	&	0.380	&	0.883	&	31.93	&		\\
$f_{8}$	&	192.435	&	0.348	&	0.806	&	26.94	&		\\
$f_{9}$	&	199.191	&	0.055	&	0.532	&	4.40	&		\\
$f_{10}$	&	203.203	&	0.064	&	0.124	&	5.04	&		\\
$f_{11}$	&	207.416	&	0.091	&	0.811	&	7.27	&		\\
$f_{12}$	&	208.769	&	0.456	&	0.706	&	36.22	&		\\
$f_{13}$	&	227.732	&	0.060	&	0.349	&	5.65	&		\\
$f_{14}$	&	230.367	&	0.186	&	0.872	&	17.93	&		\\
$f_{15}$	&	231.915	&	0.109	&	0.013	&	10.84	&		\\
$f_{16}$	&	256.997	&	0.067	&	0.862	&	6.11	&		\\
\hline   \hline                        % inserts double horizontal lines
\end{tabular}\\
\label{Tab2}
\end{table*}

%Figure 3

\subsection{Regular frequency spacing}
\label{sec4.1}
There are several well-developed methods to search for regular spacing in $\delta$ Sct stars, including the histogram graph \citep{Bre99}, the Fourier transform method \citep{Gar09, Gar13, Gar15}, the echelle diagram \citep{Pap16a, Pap16b}, and the rotational splitting analysis \citep{Che16,Che17a,Che17b}. There is no obvious and symmetric splitting multiplets for \target. Therefore, we use other three methods to study the regular spacing. Details will be described as follows. 

%Figure 4
\begin{figure}[ht!]
\begin{center}
\includegraphics[width=0.495\textwidth]{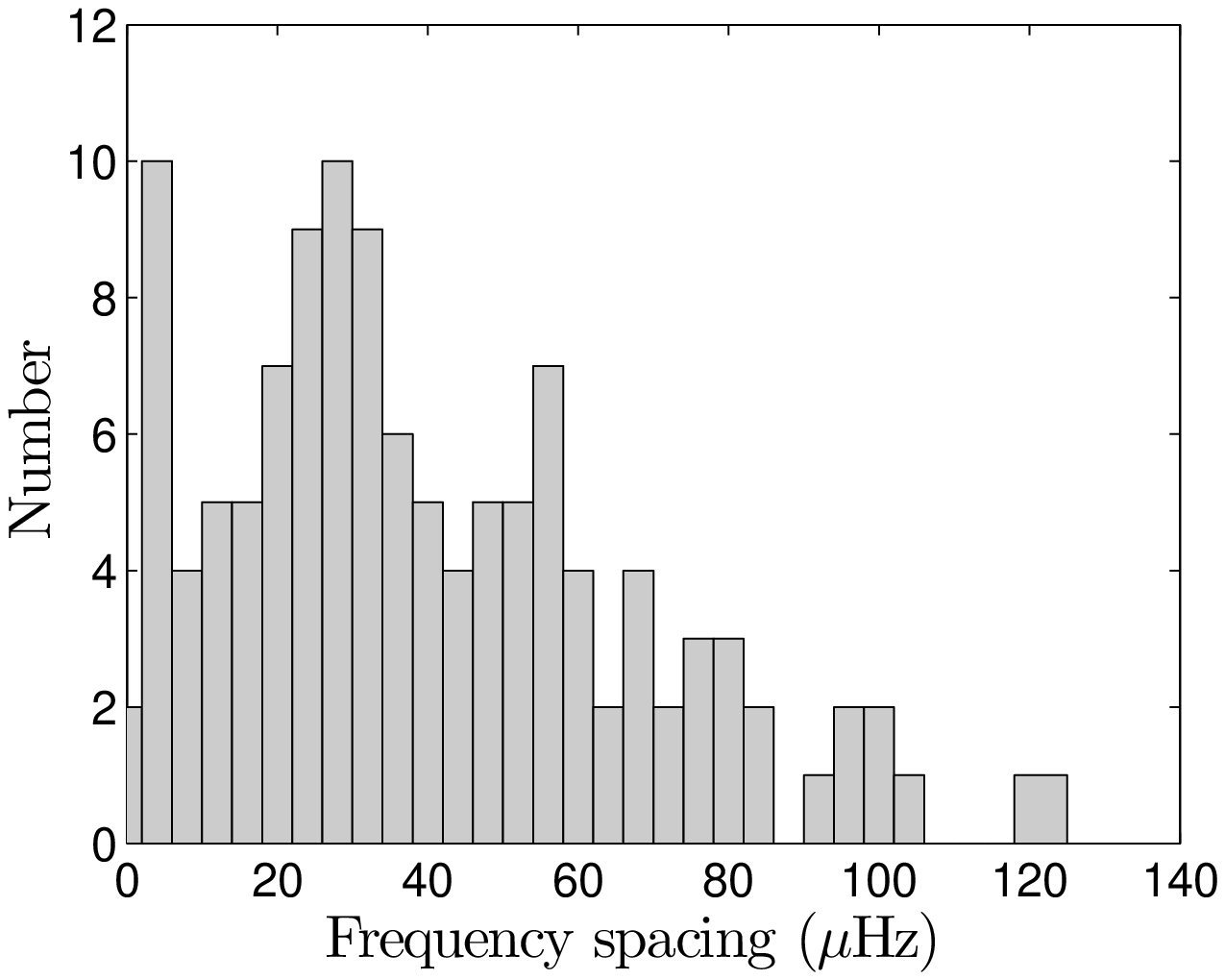}
\includegraphics[width=0.495\textwidth]{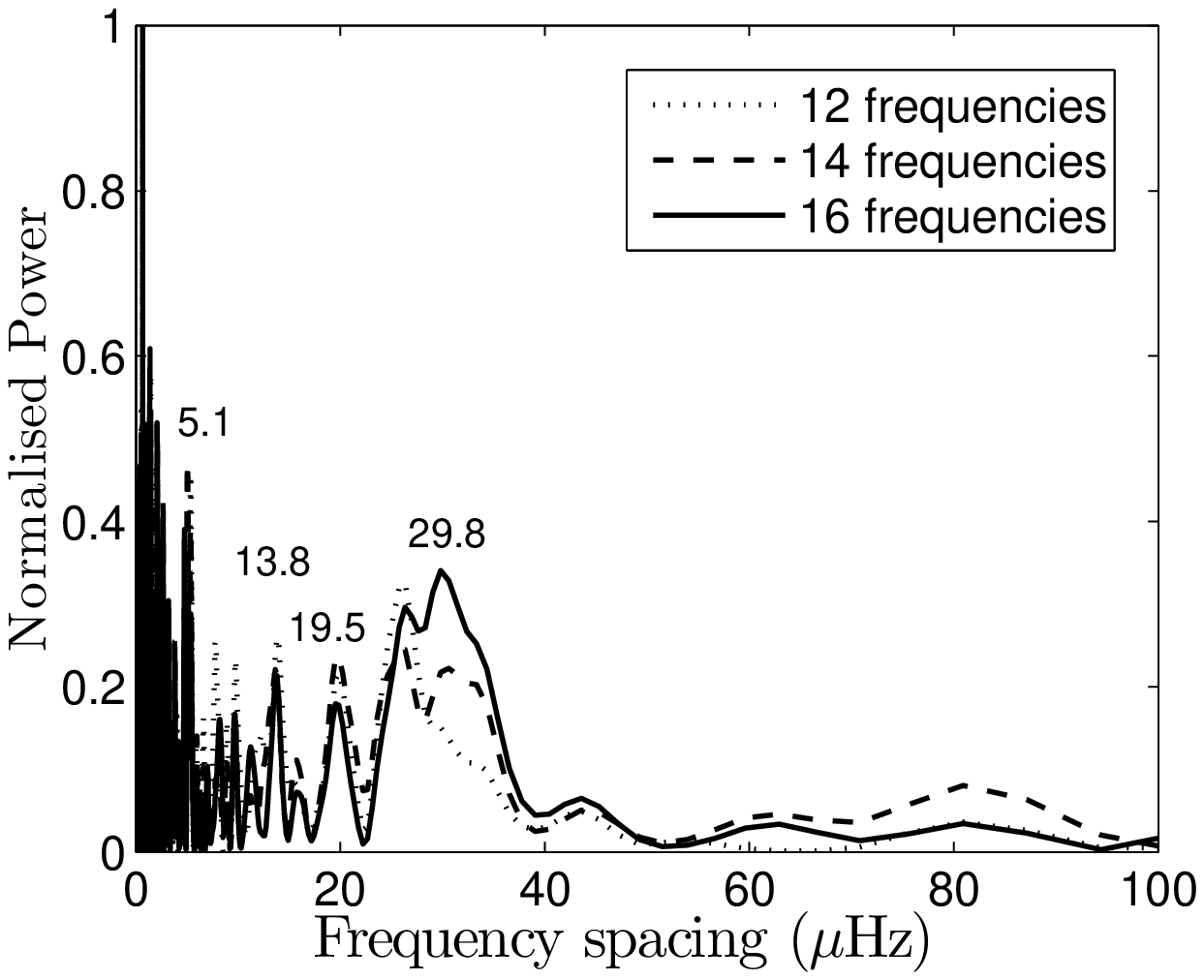}
\caption{Left panel: Distribution of the frequency differences between every two of the pulsation frequencies. Right panel: The Fourier transform of the 16 pulsation frequencies (solid line).
The Fourier transform with the highest 12 (dotted line) and 14 (dashed line) frequencies are also shown as comparisons. Detailed analyses of the regular frequency spacing are described in section \ref{sec4.1}. }
\label{Fig4}
\end{center}
\end{figure}
\subsubsection{The histogram graph}
\label{sec4.1.1}
The frequency differences between every two of the 16 extracted frequencies are calculated. The left panel of Figure \ref{Fig4} illustrates the distribution of the frequency differences. The bin size is 4\uhz. The histogram graph can be used to examine the regular frequency spacing. For \target, there are two obvious peaks around 4\uhz and 28\uhz, respectively. The first peak may be caused by close frequencies and potential rotational splitting frequencies. The significant second peak can be explained by the large separation which is caused by frequencies with the same degree $l$ of different radial orders.

%\subsection{The Fourier transform method}
\subsubsection{The Fourier transform}
The Fourier transform is another method widely used to find the regular spacing. We follow the descriptions of \cite{Gar09} to derive the frequency spacing for \target. The extracted frequencies are considered as a series of Dirac's Delta Function with equal amplitudes. Then the Fourier transform of these frequencies are calculated. We also give the Fourier transform with the highest 12 and 14 frequencies as comparisons as shown in the right panel of Figure \ref{Fig4}. A periodic pattern is recognized with all the 16 frequencies. Clear peaks appear around 5.1, 13.8, 19.5, and 29.8\uhz. Compared with the result from the histogram method (section \ref{sec4.1.1}). It indicates that probably there is a regular frequency spacing around 29.8\uhz.

%\subsection{The Echelle diagram} %ref: kic 4544587
\subsubsection{The echelle diagram}
\begin{figure}[ht!]
\begin{center}
\includegraphics[width=0.80\textwidth]{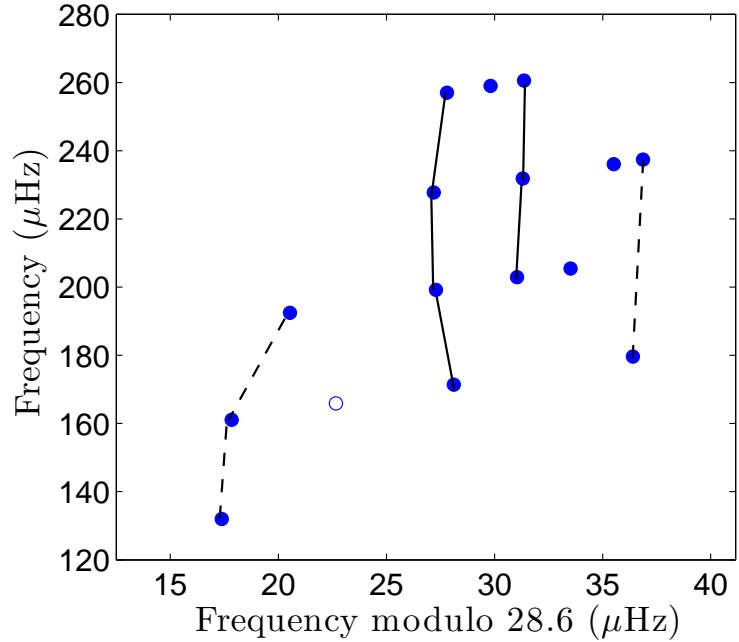}
\caption{The echelle diagram of the pulsation frequencies. The large frequency separation is chose as 28.6\uhz. There are two obvious vertical ridges (solid line), and two potential vertical ridges (dashed line).}
\label{Fig5}
\end{center}
\end{figure}

An echelle diagram \citep{Gre83, Bed10, Mos12} can directly show the regular patterns of the frequency spectrum. The histogram and Fourier transform method indicate the best regular spacing ranges from 28\uhz to 29.8\uhz. Therefore, the pulsation frequencies are modulated by a series of values between 28\uhz and 29.8\uhz. When the spacing is equal to around 28.6\uhz, the echelle diagram arranges regularly. As seen in Figure \ref{Fig5} there are two obvious vertical ridges in the echelle diagram. \cite{Ham13} speculate that the frequencies in a vertical ridge are the coupling of the self excited frequencies and the tidally induced frequencies. \cite{Lia17} proposed that there is a threshold in the binary period of about 13 days. The pulsation properties can be affected for short-period binaries because of tidal interactions. The orbital period of \target is 4.398 days, which is below the 13-day threshold. Therefore, the tidal effects may play an important role to the pulsation frequency for \target.

%Figure 5

\subsection{The mean stellar density}
\label{sec4.2}
Recently, \cite{Sua14} theoretically predicted that the mean stellar density is proportional to the large separation of $\delta$ Sct stars. \cite{Gar15} confirmed this scaling relation by considering a few eclipsing binary systems with a $\delta$ Sct component. \cite{Str18} examined that the large separation-mean density ($\Delta\nu-\rho$) relation is also suitable for semi-detached binaries. \cite{Gar17} refined this relation by implementing a Hierarchical Bayesian linear regression method. The updated relation is:
\begin{equation}
\overline{\rho}/\overline{\rho}_{\odot}=1.50^{+0.09}_{-0.10}(\Delta\nu/\Delta\nu_{\odot})^{2.04^{+0.04}_{-0.04}},
\label{Equ1}
\end{equation}
where $\Delta\nu_{\odot} = 134.8$\uhz \citep{Kje08}. Using this formula and the large separation value 28.6\uhz, the mean density is $0.09$ g$\cdot$cm$^{-3}$ for the $\delta$ Sct component of \target.

%%%%%%%%%%%%%%%%%%%%%%
%%%%%%%% Section 6  %%%%%%%%
%%%%%%%%%%%%%%%%%%%%%%
\section{Results and conclusions}\label{sec5}
In this paper, we confirm that \target is a short-period eclipsing binary containing a $\delta$ Sct variable. \kep has observed \target for about 77 days. The high-precision light curve is suitable for binary modelling and pulsation analysing. The binary fitting results reveal that the two stars have similar radius. This system may have experienced orbital circularization because the orbital eccentricity is very close to zero. 

After removing the eclipsing variations, the residual light curve shows 16 pulsating frequencies between 120\uhz and 270\uhz with a signal-to-noise ratio larger than 4. The dominant peak has a pulsation amplitude of $\sim$0.5 mmag at $f_{12}=208.769$\uhz. A regular frequency spacing is cross-identified with the histogram graph, the Fourier transform, and the echelle diagram method. The regular spacing is about 28.6\uhz, which is approximately 11 times of the orbital frequency (28.95\uhz). It implies that the pulsations of the $\delta$ Sct star may have been influenced by the companion's tidal forces. Using the scale relation between the mean stellar density and the large frequency separation, we obtained a mean density of $0.09$ g$\cdot$cm$^{-3}$ for the $\delta$ Sct component of \target.

Although \kep has accomplished its mission, there are still a large amount of data waiting to be analysed. \target was first recognized as an exoplanet, but then corrected to be an eclipsing binary system. In our work, by analyzing the \kep data we are able to confirm that this system contains a $\delta$ Sct component. In the future, more systems like EPIC 202843107 can help scientists study the differences in evolutionary and physical status between single $\delta$ Sct stars and those in binary systems.

\begin{acknowledgements}
This work is supported by the National Natural Science Foundation of China (Grant No. 11803012,11333002,11673011,11503009), and the National Defense Science and Engineering Bureau civil space flight advanced research project (D030201). 
\end{acknowledgements}

\label{lastpage}

\end{document}